\documentclass[runningheads,orivec]{llncs}

\usepackage[T1]{fontenc}
\usepackage{graphicx}
\usepackage{booktabs}
\usepackage{amsmath}
\usepackage{isomath}
\usepackage{color,soul}
\usepackage{microtype}

\pdfoutput=1

\usepackage{graphicx}
\usepackage{ifpdf}
\ifpdf
\DeclareGraphicsExtensions{.pdf,.ai,.jpg,.png}
\setkeys{Gin}{pagebox=artbox}
\else
\DeclareGraphicsExtensions{.ai,.ps,.eps}
\fi
\graphicspath{{./}{./ecir19-wikipedia-text-reuse-paper-figures/}}

\newif\ifbscomment
\bscommentfalse
\bscommenttrue
\RequirePackage{ifthen}
\RequirePackage{type1cm}
\RequirePackage{xcolor}
\RequirePackage{soul}
\setstcolor{blue}

\colorlet{exampleHlColor}{green!15!lightgray!60}
\sethlcolor{exampleHlColor}

\newcommand{\Ni}{(1)~}
\newcommand{\Nii}{(2)~}

\newcommand{\Na}{(a)~}
\newcommand{\Nb}{(b)~}
\newcommand{\Nc}{(c)~}
\newcommand{\Nd}{(d)~}

\usepackage{xspace}

\newcommand{\tfidf}{\ensuremath{\mathit{tf}\kern-0.15em\cdot\kern-0.15em\mathit{idf}}\xspace}
\newcommand{\tfidfBf}{\ensuremath{\mathbfit{tf}\kern-0.15em\cdot\kern-0.15em\mathbfit{idf}}\xspace}

\begin{document}

\title{Wikipedia Text Reuse: Within and Without}

\author{%
Milad Alshomary,\textsuperscript{1}
Michael Völske,\textsuperscript{2}
Tristan Licht,\textsuperscript{2}
Henning Wachsmuth,\textsuperscript{1}\\[1ex]
Benno Stein,\textsuperscript{2}
Matthias Hagen,\textsuperscript{3}
and 
Martin Potthast\textsuperscript{4}}

\authorrunning{Alshomary et al.}

\institute{%
\textsuperscript{1}Paderborn University\\
\textsuperscript{2}Bauhaus-Universität Weimar\\
\textsuperscript{3}Martin-Luther-Universität Halle-Wittenberg\\
\textsuperscript{4}Leipzig University}

\maketitle

\begin{abstract}
	We study text reuse related to Wikipedia at scale by compiling the first corpus of text reuse cases within Wikipedia as well as without (i.e., reuse of Wikipedia text in a sample of the Common Crawl). To discover reuse beyond verbatim copy and paste, we employ state-of-the-art text reuse detection technology, scaling it for the first time to process the entire Wikipedia as part of a distributed retrieval pipeline. We further report on a pilot analysis of the 100~million reuse cases inside, and the 1.6~million reuse cases outside Wikipedia that we discovered. Text reuse inside Wikipedia gives rise to new tasks such as article template induction, fixing quality flaws due to inconsistencies arising from asynchronous editing of reused passages, or complementing Wikipedia's ontology. Text reuse outside Wikipedia yields a tangible metric for the emerging field of quantifying Wikipedia's influence on the web. To foster future research into these tasks, and for reproducibility's sake, the Wikipedia text reuse corpus and the retrieval pipeline are made freely available.
\end{abstract}

\section{Introduction}

Text reuse is second nature to Wikipedia: {\em inside} Wikipedia, the articles grouped in a given category are often harmonized until templates emerge, which are then adopted for newly created articles in the same category. Moreover, passages are copied verbatim from one article to another, such as, in cases where they form a hierarchical relationship. While the reuse of text inside Wikipedia has been a de facto policy for many years, neither the MediaWiki software nor tools developed by and for the Wikipedia community offer any reuse support beyond templating: unless a dedicated Wikipedia editor takes care of it, a copied passage will eventually diverge from its original, resulting in inconsistency.

{\em Outside} Wikipedia, we distinguish reuse of Wikipedia's articles by third parties, and reuse of third-party content by Wikipedia. The former is widespread: for example, passages of articles are manually reused in quotations and summaries, or automatically extracted to search result pages. Many sites mirror Wikipedia partially or in full---sometimes with proper attribution, other times violating Wikipedia's lenient copyrights. The latter form of reuse is discouraged.

With a few exceptions noted below, Wikipedia text reuse has not been analyzed at scale. We attribute the lack of studies to the lack of open and scalable technologies capable of detecting text reuse regardless of reformatting, paraphrasing, and light summarization, as well as to the significant computational overhead required to process the entire Wikipedia. Only recently, resulting from a six-year effort to systematically evaluate reuse detection algorithms in the PAN~lab at the CLEF~conference, new classes of algorithms emerged that specifically address the detection of such kinds of text reuse from large text corpora.

To foster research into Wikipedia text reuse, we compile the first Wikipedia text reuse corpus, obtained from comparing the entire Wikipedia to itself as well as to a 10\%-sample of the Common Crawl. For this purpose, we scaled up the aforementioned algorithms for the first time to render the computations feasible on a mid-sized cluster. Finally, we carry out a first exploratory analysis, enabling us to report on findings that cover insights on how Wikipedia text reuse detection could be used as a tool for both helping the Wikipedia community to ensure the encyclopedia's consistency and coherence, as well as for quantifying the influence of Wikipedia on the web at large.

\section{Related Work}

Wikipedia's openness and success fuels tons of research about the encyclopedia%
\footnote{\url{https://en.wikipedia.org/wiki/Academic_studies_about_Wikipedia}}
as well as on how it can be exploited, e.g., for disease forecasting~\cite{generous:2014}, movie success prediction~\cite{mestyan:2013}, and much more. Wikipedia's influence on the web has recently become a focus of interest: for instance, posts on Stack Overflow and Reddit that link to Wikipedia have been found to outperform others in terms of interactions~\cite{vincent:2018}. Other works have studied Wikipedia's role in driving research in the scientific community~\cite{thompson:2018}, and its importance to search engines in enriching search result pages~\cite{mcmahon:2017}. The ever increasing quality of Wikipedia drives the reuse of its content by third parties, but in a ``paradox of reuse'' reduces the need to visit Wikipedia itself~\cite{taraborelli:2015}, depriving the encyclopedia of potential new editors.

In general, text reuse detection is applied in many domains~\cite{bendersky:2009c}, such as the digital humanities (to uncover the influence of ancient authors~\cite{coffee:2012}), and in journalism and science (to study author perspectives~\cite{clough:2001} as well as to pursue copyright infringement and plagiarism~\cite{citron:2015}). Text reuse detection divides into the subtasks of \emph{source retrieval} and \emph{text alignment}~\cite{stein:2013h,stein:2007f}, where the former retrieves a set of candidate sources for reuse given a questioned document, and the latter the mutually reusing passages given a document pair. Approaches addressing each task have been systematically evaluated at PAN~\cite{stein:2013h}.

Regarding Wikipedia in particular, text reuse detection has the potential to aid the community in improving the encyclopedia as well to serve as a new method of quantifying its influence on the web. However, Wikipedia text reuse has been mostly disregarded to date, except for two pioneering studies: Weissman et~al.~\cite{weissman:2015} report on text reuse within Wikipedia, and its connection to article quality. Using similarity hashing, near-duplicate sentences are identified that are redundant or contradictory. Similarly, Ardi et~al.~\cite{ardi:2014} employ hashing to detect near-duplicates of Wikipedia articles in the Common Crawl. Both neglect the text alignment step, yielding reuse cases that are either too fine-grained or too coarse-grained for reuse analysis. Our text reuse detection pipeline incorporates similar hashing techniques for source retrieval but further filters and refines the results through text alignment to obtain the actual reused text passages. In this respect, our corpus captures better the original user intent of reusing a given passage of text while saving its users the computational overhead.

\section{Corpus Construction}

Given two document collections $D_1$ and $D_2$, we aim to identify all cases of text reuse as pairs of sufficiently similar text spans. Here, $D_1$ comprises all English Wikipedia articles and $D_2=D_1$ for within-Wikipedia detection, whereas otherwise $D_2$ comprises a 10\%~sample of the Common Crawl (see Table~\ref{table-dataset-statistics} (left)). Our processing pipeline first carries out {\em source retrieval} to identify promising candidate document pairs, which are then compared in detail during {\em text alignment}.

\subsection{Source Retrieval}

In source retrieval, given a questioned document $d_1\in D_1$ and $D_2$, the task is to rank the documents in $D_2$ in order of decreasing likelihood of sharing reused text with~$d_1$. An absolute cutoff rank~$k$ and/or a relative score threshold~$\tau$ may be used to decide how many of the top-ranked documents become subject to the more expensive task of text alignment with~$d_1$. These parameters are typically determined in terms of the budget of computational capacity available as well as the desired recall-level. An ideal ranking function would rank all documents in $D_2$ that reuse text from $d_1$ highest; however, the typical operationalization using text similarity measures does not reach this ideal. The higher the desired recall-level, the lower the precision and the higher the computational overhead. 

Our computational budget were 2~months worth of processing time on a 130~node cluster (12~CPU cores and 196~GB RAM each) running Apache Spark, an the goal was to maximize recall. Since $D_1$~(Wikipedia) as a whole is questioned, we generalized source retrieval toward a pruned all-pairs similarity search, ranking all $(d_1,d_2)\in D_1\times D_2$ based on the following pruned ranking function~$\rho$:
$$
\underbrace{\exists c_i\in d_1, c_j\in d_2\!:\quad h(c_i) \cap h(c_j)\neq\emptyset}_{\mbox{\scriptsize Search pruning}}\quad \to\quad  \rho(d_1, d_2) = \max_{\substack{c_i \in d_1\\c_j \in d_2}}(\varphi(c_i, c_j)),
$$
where $c$~is a passage-length chunk of document~$d$, $h$~is a locality-preserving hash function, and $\varphi$ a text similarity measure. The design goal was to apply great leniency during search pruning (a single hash collision suffices for rank scoring), and to take into account that reuse is a passage-level phenomenon.

To decide upon the hash function~$h$ and the similarity measure~$\varphi$, and to tune their parameters, we compiled a ground truth by applying the subsequent step of text alignment directly, extracting the mutual reuse of each of 1000~long Wikipedia articles from all other articles. This way, $h$, $\varphi$, and hence~$\rho$ are optimized against our text alignment algorithm so as not to prevent its detections. We considered two hashing schemes for~$h$:
\Ni
random projections, an instantiation of the data-independent locality-sensitive hashing (LSH) family~\cite{charikar:2002}, and
\Nii
variational deep semantic hashing (VDSH), a data-dependent learning-to-hash technique~\cite{chaidaroon:2017}.
We further considered four text similarity measures for~$\varphi$:
\Na
cosine similarity on a \tfidf-weighted unigram representation,
\Nb
Jaccard similarity on stop word $n$-grams~\cite{stamatatos:2011},
\Nc
cosine similarity on a simple additive paragraph vector model~\cite{mitchell:2008}, and
\Nd
a weighted average of~(b) and~(c).

Table~\ref{table-dataset-statistics} (middle)  shows our evaluation results for the two components of the source retrieval pipeline. For search pruning, we selected VDSH with a 16-bit hash, which reduces the number of required evaluations of the $\rho$ measure by three orders of magnitude compared to an exhaustive comparison, while retaining the majority of text reuse cases. To construct the ranking function $\rho$ itself, we settle on cosine similarity in the \tfidf space as the similarity measure~$\varphi$ for its superior recall compared to all other models.

\begin{table*}[t]
	\centering
	\scriptsize
	\caption{Overview of input datasets (left), source retrieval performance (middle), and retrieved text reuse (right). Counts in the tables left and right are in millions.}
	\label{table-dataset-statistics}
	\vspace{-3ex}
	\renewcommand{\arraystretch}{1.2}%
	\begin{tabular}[t]{@{}lr@{}}
		\toprule
		\bfseries Dataset & \bfseries Count             \\
		\midrule
		\multicolumn{2}{@{}l@{}}{\itshape Wikipedia}    \\
		Articles          & 4.2                         \\
		Paragraphs        & 11.4                        \\
		\midrule
		\multicolumn{2}{@{}l@{}}{\itshape Common Crawl} \\
		Websites          & 1.4                         \\
		Web pages         & 591.0                       \\
		Paragraphs        & 187.0                       \\
		\bottomrule
	\end{tabular}%
	\hfill%
	\renewcommand{\arraystretch}{0.78}%
	\begin{tabular}[t]{@{}l@{\quad}r@{\quad}r@{}}
		\toprule
		\bfseries Source Retrieval & \bfseries Recall & \bfseries Precision \\
		\midrule
		\multicolumn{3}{@{}l@{}}{\itshape Search pruning}                   \\
		\Ni LSH                & 0.32           & 9.8$\cdot 10^{-6}$           \\ 
		\bfseries \Nii VDSH    & \bfseries 0.73 & \bfseries 4.5$\mathbf{\cdot 10^{-4}}$ \\
		\midrule
		\multicolumn{3}{@{}l@{}}{\itshape Ranking up to rank $k=1000$} \\
		\bfseries \Na \tfidfBf  & \bfseries 0.87 & \bfseries 0.007      \\ 
		\Nb Stop $n$-grams      & 0.74           & 0.007                \\
		\Nc Par2vec             & 0.67           & 0.008                \\
		\Nd Hybrid              & 0.76           & 0.009                \\
		\midrule
		\bfseries VDSH + \tfidfBf & \bfseries 0.66 & \bfseries 0.005      \\
		\bottomrule
	\end{tabular}%
	\hfill%
	\renewcommand{\arraystretch}{1}%
	\begin{tabular}[t]{@{}l@{\quad}r@{\quad}r@{}}
		\toprule
		\bfseries Reuse & \bfseries Within & \bfseries Without        \\
		\midrule
		Cases           & 110              & 1.6                      \\
		\midrule
		\multicolumn{3}{@{}l@{}}{\itshape Documents with Reuse Cases} \\
		Articles        & 0.360            & 1.0                      \\
		Pages           & --               & 0.015                    \\
		\midrule
		\multicolumn{3}{@{}l@{}}{\itshape Words in Reuse Cases}       \\
		Min.            & 17               & 23                       \\
		Avg.            & 78               & 252                      \\
		Max.            & 6\,200           & 1960                     \\
		\bottomrule
	\end{tabular}%
	\vspace*{-4ex}%
\end{table*}

\subsection{Text Alignment}

Given a candidate document pair, text alignment extracts spans of reused text---if any---through the steps \textit{seed generation} (identification of short exact matches), \textit{seed extension} (clustering of matches to form spans), and \textit{post filtering}. The state of the art evaluated at PAN is determined on datasets orders of magnitude smaller than our setting, often using complex setups that turned out to be difficult to scale and to be reproduced, while lacking open source implementations. We hence resorted to ideas from the literature that offer a reasonable trade-off between performance, robustness, and speed, and tuned their parameters%
\footnote{We employed word 3-gram seeds, extended via DBScan clustering ($\varepsilon=150$ and $\mathrm{minPoints}=5$), and filtered cases less than 200 words long or 0.5~cosine similarity.}
based on the standard PAN-13 training data. Our text alignment achieves a macro-averaged \emph{plagdet} score of~0.64 (0.84 on just the unobfuscated subset) on the corresponding PAN-13 test data. In terms of raw detection performance, this is in the lower middle range of PAN results~\cite{stein:2014k}.

In our pruned all-pairs search setting, each input record to the text alignment step specifies the candidate pairs involving one document $d_i$ from collection $D_1$, in the form of a list of all candidate documents from the other collection $D_2$ sorted by descending $\rho$-score. Text alignment is applied sequentially to this list until one of two stopping criteria is met: \Ni the current candidate pair is below the \emph{$\rho$-threshold} (0.025 in our implementation), or \Nii another threshold on the \emph{number of consecutive miss-cases}---i.e., candidate pairs in which the text alignment finds no reuse---is exceeded (we use 250). Both thresholds can be configured based on the time available for the task; our values were determined experimentally using the aforementioned ground-truth sample.

\section{Corpus Analysis}

Table~\ref{table-dataset-statistics} (right) shows basic statistics of the text reuse we uncovered; most interestingly, we find nearly~70 times more reuse cases within Wikipedia than without, but involving only one third as many articles. Based on this insight, we identify two fundamentally different kinds of text reuse within Wikipedia, the first of which makes up for the bulk of this discrepancy. When articles use the same text structure, but different facts---e.g., when geographical locations are described in terms of their surroundings---we refer to this as \emph{structure reuse} (Table~\ref{table-text-reuse-examples}, top left) and consider it non-problematic, and perhaps unavoidable redundancy. Resulting from Wikipedia's editorial process, structure reuse is much more likely to occur within Wikipedia than without. On the other hand, articles may contain factually nearly-identical passages, likely after copying from one to the other. We consider such \emph{content reuse} likely to result in inconsistency and contradiction as the articles diverge over time (Table~\ref{table-text-reuse-examples}, bottom left). Ideally, such redundant sections should be replaced with a single, authoritative source. In this sense, text reuse analysis can help the Wikipedia community locate and improve articles with undesirable redundancy.

Our current observations indicate that the ontological relationship between articles' topics correlates with the type of text reuse: Structure reuse occurs more frequently when articles represent concepts on the same level in the ontology tree (Table~\ref{table-text-reuse-examples}, top right), while two articles whose subjects are vertically aligned---as is the case with ``is a'' or ``part of'' relationships---are more likely to exhibit content reuse (Table~\ref{table-text-reuse-examples}, bottom right). The latter association can also be envisioned as a solution to the sub-article matching task~\cite{lin:2017}: the occurrence of content reuse between articles can serve as an indicator of the ontological relationship between the concepts that they represent. However, distinguishing content and structure reuse automatically is not trivial. Our initial attempt at classifying reuse cases used heuristics based on the ratio of reused to original text in the articles, as well as the Jaccard similarities between the sets of named entities and word 10-grams. Based on two samples of 100 random cases classified as each, this yields 100\% precision for structure reuse, but only 57\% for content reuse. While our heuristics identify 95.5 million~(87\%) of all within-Wikipedia cases as structure reuse, the true number likely exceeds 100 million assuming our error estimates are accurate.

\newcommand{\exampleEllipsis}{{\tiny\color{gray}[\small\hspace{.15em}\dots\tiny]}}
\begin{table*}[tb]
	\caption{Examples of the two types of text reuse within Wikipedia---structure reuse (top) and content reuse (bottom)---along ontological article relations (right).}
	\label{table-text-reuse-examples}
	\vspace{-3ex}
	\begin{minipage}[t]{0.74\textwidth}
		\scriptsize
		\begin{tabular}[t]{@{}p{0.47\linewidth}@{\hspace{0.025\linewidth}}p{0.47\linewidth}@{}}
			\toprule
			{\bf Title:} Nied\'{z}wiedzie, Pisz County &
			{\bf Title:} Zimna Woda, Zgierz County \\
			\midrule
			Nied\'{z}wiedzie \hl{is a village in the administrative district of Gmina} Pisz, within Pisz County, Warmian-Masurian \hl{Voivodeship, in} northern \hl{Poland. It lies approximately} south-east of Pisz and east \hl{of the regional capital}
			Olsztyn.
			&
			Zimna Woda \hl{is a village in the administrative district of Gmina} Zgierz, within Zgierz County, \L{}\'od\'z \hl{Voi\-vodeship, in} central \hl{Poland. It lies approximately} north-west of Zgierz and north-west \hl{of the regional capital} \L{}\'od\'z.
			\\
			\bottomrule
		\end{tabular}\\[1ex]
		\begin{tabular}[t]{@{}p{0.47\linewidth}@{\hspace{0.025\linewidth}}p{0.47\linewidth}@{}}
			\toprule
			{\bf Title:} Human tooth development&
			{\bf Title:} Tooth eruption\\
			\midrule
			\hspace{3em} Tooth eruption has \hl{three stages. The first, known as} deciduous \hl{dentition stage, occurs when only primary teeth are visible. Once the first permanent tooth erupts into the mouth, the teeth are in the mixed (or transitional) dentition}.
			\exampleEllipsis\par
			\hl{Primary dentition} stage \hl{starts on the arrival of the mandibular central incisors,} typically from around six \hl{months, and lasts until the first permanent molars appear} 
			\exampleEllipsis
			&
			The dentition goes through \hl{three stages. The first, known as} primary \hl{dentition stage, occurs when only primary teeth are visible. Once the first permanent tooth erupts into the mouth, the teeth} that are visible \hl{are in the mixed (or transitional) dentition} stage.
			\exampleEllipsis\par
			\hl{Primary dentition starts on the arrival of the madibular central incisors,} usually at eight \hl{months, and lasts until the first permanent molars appear} 
			\exampleEllipsis
			\\
			\bottomrule
		\end{tabular}
	\end{minipage}%
	\hspace{0.5em}%
	\begin{minipage}[t]{0.24\textwidth}
		\vspace*{0ex}
		\includegraphics[width=\textwidth]{ontology-reuse1}\\[6ex]
		\includegraphics[width=\textwidth]{ontology-reuse2}
	\end{minipage}%
	\vspace*{-4ex}
\end{table*}

\enlargethispage{\baselineskip}
In the Common Crawl sample we examined, 4,898~websites host at least one page that reuses text from a Wikipedia article for a total of 1.6~million cases.%
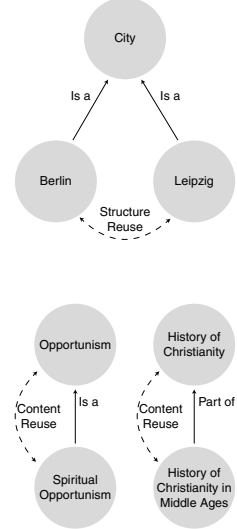
\footnote{The top three being \url{wikia.com} (563), \url{rediff.com} (55), and \url{un.org} (28~reusing pages).}
We presume that Wikipedia's policy of avoiding reuse from third parties inside its articles is enforced by its editors, so that nearly all of the cases will be third parties reusing Wikipedia's articles instead. Most (94\%) of the pages violate the terms of Wikipedia's license%
\footnote{\url{https://en.wikipedia.org/wiki/Wikipedia:Reusing_Wikipedia_content}}
by not referencing Wikipedia as a source (i.e., the term ``Wikipedia'' does not occur). With only a handful exceptions, such as un.org, all of the sites display advertisements, which extends to the pages containing the reuse. Furthermore, in nearly all of the cases, the reuse accounted for more than~90\% of the main content, begging the question of their usefulness.

To quantify the degree to which sites reusing Wikipedia content may be profiting from this reuse, we {\em conservatively} estimate the potential advertisement revenue generated by the reused content. We make the simplifying assumption that all reusing websites host only one ad per page and that advertisements are billed according to cost per mille (CPM), achieving a revenue per mille (RPM) of about half~(1.4~USD) the average estimated CPM on the web in 2018~(2.8~USD).%
\footnote{\url{https://monetizepros.com/cpm-rate-guide/display/}}
Accounting for the fact that reusing pages are generally ranked lower than Wikipedia in search results, we use 10\%~of the monthly page view counts on reused articles (as per Wikipedia's~API) as estimates for the page views of reusing pages. With these approximations, we arrive at a {\em lower bound estimate} of 45,000~USD monthly ad revenue generated by these 4,898~sites. Extrapolated to the entire web (say, 600,000 reusing sites out of 180~million active sites as per netcraft.com), we arrive at 5.5 million~USD estimated monthly ad revenue; which adds up to about 72\% of Wikipedia's last year's worldwide fundraising returns.

\enlargethispage{2\baselineskip}
\vspace*{-2ex}

\section{Conclusion}
\vspace{-1ex}

In an effort to bring text reuse analysis to very large corpora, we propose a scalable pipeline comprising the \textit{source retrieval} and \textit{text alignment} subtasks. We address challenges of scale primarily in the former by way of candidate filtering, and evaluate a set of hashing and text similarity techniques for this purpose. Using our framework, we compile two text reuse datasets---\textit{within} and \textit{without} Wikipedia---which we make publicly available for further research into text reuse phenomena. We hope our data will stimulate future research targeting Wikipedia quality improvement---e.g., by template induction or automatic detection of inconsistency---and understanding Wikipedia's influence on the web at large.

\begin{raggedright}

\end{raggedright}
\end{document}